# Electron-Ion Collider with Quasi-Ordered Ion Beam

I. Meshkov[*]

*Joint Institute for Nuclear Research, Joliot-Curie, 6, Dubna 141980, Russia*
*St.Petersburg State University, University Emb., 7–9, St.Petersburg 199034, Russia*

**Abstract**

The possibility of using in an electron-ion collider the so-called quasi-ordered ion beam deep cooled by electron cooling is considered. The properties of the ordered and quasi-ordered beams are described. The advantages of using the quasi-ordered beam are discussed.

## 1. Introduction

One of the most common methods for studying the structure of rare and exotic nuclei is electron-ion scattering. A long-discussed version of this method is its implementation in the collider mode. However, the use of the collider mode is limited by a low intensity of the beams of exotic nuclei (ions).

In this paper we consider the possibility of using the so-called quasi-ordered deep-cooled (by electron cooling) ion beam in an electron-ion collider. These experimentally discovered beams have their own fascinating history.

Shortly after observing the tremendous suppression of the Schottky noise at a decrease in the intensity of the deep cooled proton beam on the NAP-M storage ring [1], V. Parkhomchuk formulated the idea of crystalline beams [2]. This aroused great interest among accelerator physicists, which was followed by experiments on several cooler storage rings constructed and commissioned in the beginning 1990s. Then, the physics of the beam "crystallization" effect in storage rings with electron cooling was thoroughly studied experimentally and analyzed theoretically.

In 1996, a qualitatively new behavior of the ion beam was discovered by M. Steck et al. on the ESR storage ring [3] The ion beam, with a decrease in its intensity up to certain value, suddenly undergoes an abrupt "jump downward" (!) in the values of its parameters — the ion momentum spread and the transverse dimensions (Fig. 1a) — and both parameters remain constant with further decrease of the intensity. This was the *phase transition* to the state of the crystalline beam. As was soon realized, such a crystal was a one-dimensional chain of ions. Therefore, such a beam was called "*ordered*".

In subsequent years, ordered beams were obtained for a wide variety of ions of a wide energy range, from carbon $C^{6+}$ to uranium $U^{92+}$ in storage rings ESR [4], SIS-18 [5] (both GSI) and CRYRING (Stockholm university) [6], S-LSR (Kyoto University) [7] (Fig. 1). The "enigma" of the NAP-M experiment, where no phase transition was observed in the deep cooled proton beam, was solved in the end. At the third attempt, after the failures at COSY (FZ Juelich) and ESR, the phase transition of the proton beam in the S-LSR storage ring (Kyoto University) was obtained (Fig. 1c) [8]. By common opinion, the two previous failures were caused by the insufficient stability ("ripples") of the dipole magnetic field of the storage rings, which led to variation in the particle orbit length and limited stability of high voltage power supply of the electron cooler.

---

[*] e-mail: meshkov@jinr.ru

Soon after successful experiments on the ordered ion beam formation the first proposals of colliders with ordered ion beams appeared [10,11]. However, luminosity of these colliders is substantially limited by the low intensity of the ordered ion beams. Several proposals on how to increase the intensity of ordered beams were made but not developed further. Nevertheless, all the properties of crystalline beams and, no less important, beams above the phase transition have not been used up to now.

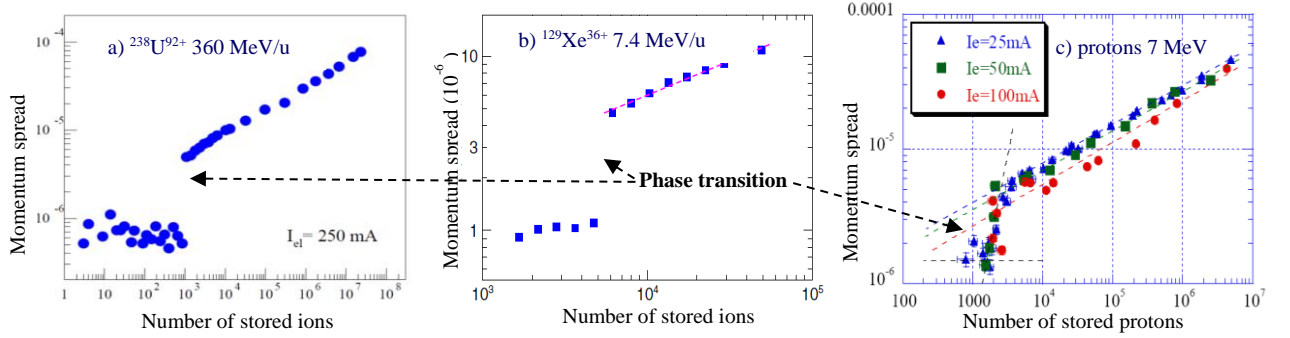

Fig. 1. Particle momentum spread (1 σ) vs particle number. (a) $^{238}U^{92+}$ beam in ESR [3]. (b) $^{129}Xe^{36+}$ in CRYRING [6]. (c) proton beam in S-LSR [8].

## 2. Characteristics of ordered ion beams below and above transition

In this paper we consider the feasibility of using such "quasi-ordered beams" in electron-ion collisions to study rare and exotic nuclei. This application is of interest for the projects ELISe (ELectron-Ion Scattering in a Storage ring) at FAIR [12] and DERICA (Dubna Electron Radioactive Ion Collider fAcility) at JINR [13].

Summarizing the properties of the ion beams above and under phase transition in all *five rings* mentioned above (Table I) one should note the following.

• The transition of the ion beam to the ordered state occurs under the action of electron cooling, when the distance between the particles is increased to a certain value

$$\Delta s \geq \Delta s_{transition}. \qquad (1)$$

• The linear density *dN/ds* of the ions at transition is extremely high in the NAP-M and very low in the ESR rings, whereas it has practically equal values in three other rings. The first one and the last have the focusing structure with periodicity that meets the ordering condition [14]:

$$2\sqrt{2} \cdot \{Q_x, Q_y\} \leq \text{Lattice periodicity}. \qquad (2)$$

Here $Q_x$, $Q_y$ are betatron tunes.

• As the ion number in the beam $N_i$ increases above the value at transition, the beam transverse size $\sigma_\perp$ and particle momentum spread grow as a cubic root of $N_{beam}$ (Fig. 1). This behavior of the beam parameters was found in experiments at the ESR [9], CRYRING [6], and S-LSR [8]. It results from the equilibrium of the particle intrabeam scattering above transition, which is inversely proportional to the square of the particle velocity, and the cooling force proportional to the velocity of the cooled particles, both in the particle rest frame. Correspondingly, the beam emittance varies with the beam ion number $N_{ion}$ as follows:

$$\varepsilon(N_i) = \varepsilon_{transition} \left(\frac{N_i}{N_{transition}}\right)^{2/3}. \tag{3}$$

Here $\varepsilon_{transition}$ and $N_{transition}$ are the ion beam parameters at transition.

The law $(N_i)^{1/3}$ continues at least up to $N_i \sim 10^7$ ($dN/ds \sim 5 \cdot 10^5$/m) for all three rings regardless of the large difference in the linear density of the beams in the ESR and two others. Further increase in the particle number is limited by the space charge effect [6,9].

TABLE I. Parameters of the ordered beams.

| Storage ring | NAP-M | ESR | | SIS-18 | CRYRING | S-LSR |
|---|---|---|---|---|---|---|
| Reference | [1] | [3,6] | | [10] | [4] | [5] |
| Parameters of the cooler storage rings | | | | | | |
| $C_{Ring}$, m | 47 | 108.4 | | 216 | 51.63 | 22.557 |
| (Super) Periodicity | 4 | 2 | | (6) 12 | 6 | 6 |
| $Q_x, Q_y$ | 1.2, 1.4 | (2.1–2.45)/(2.1–2.45) | | 4.0–4.25 | 2.42, 2.42 | 1.45, 1.44 |
| Average beta-function, m | | 22.3 | | | | 2.5 |
| Ion | proton | $^{12}C^{6+}$ | $^{70}Zn^{30+}$ $^{238}U^{92+}$ | $^{238}U^{92+}$ | $^{129}Xe^{36+}$ | proton |
| $E_{ion}$, MeV/amu | 65 | 400 | | 11.4 | 7.4 | 7.0 |
| Beam parameters at transition | | | | | | |
| $N_{transition}$ | $2 \times 10^7$ | $10^3$ | | $2 \times 10^4$ | $5 \times 10^3$ | $4 \times 10^3$ |
| $\Delta s_{transition}$, cm | $2.5 \times 10^{-4}$ | 10.8 | | 1.1 | 1.0 | 1.1 |
| $(dN/ds)_{transition}$, m$^{-1}$ | $4 \times 10^5$ | 10 | | 91 | 97 | 91 |
| $\sigma_{transition}$, mm | 0.1 | 0.04–0.2 | | — | — | 0.02 |
| Emittance, nm | 1.6 | 1.8 | | | | 0.16 |
| $(\Delta p/p)_{transition}$ (1σ), $10^{-6}$ | — | 2.0 | | 1.0 | 3.3 | 5.0 |
| Parameters of ordered beams | | | | | | |
| $(\Delta p/p)_{ordered}$ (1σ), $10^{-6}$ | 1.1 | 0.2 | | 2.0 | 1.0 | 1.03 |
| $\sigma_{ordered}$, mcm | <100 | 2.0–5.0 | | — | — | <10 |
| Emittance, nm | <1.6 | $(0.18–1.1) \times 10^{-3}$ | | — | — | ≤0.04 |

• As can be concluded from the results presented in Fig. 1c, the minimum value of the particle momentum spread (and transverse size as well) above transition slightly decreases with enhancement of the cooling electron beam current.

• The experimental results [6,9] also show that the linear density of a bunch of an ordered beam cannot exceed the density of a coasting beam, i.e., it is limited by the same condition (1). It means that bunching of ordered beams for increasing luminosity of a collider makes sense if the number of particles injected into collider is limited as well. This is the case, for instance, in the study of radioactive and/or rare isotopes.

## 3. Luminosity of electron-ion collider

Next we consider the features of an electron-ion (e-i) collider of extremely low intensity. The ion beam of the collider is proposed to be used in two modes — a coasting beam or a bunched beam.

Luminosity of the collider with a coasting ion beam and a bunched electron beam can be evaluated from the formula that differs from the well-known luminosity formula for a collider with both bunched beams (see details in Ref. [15]). In the case of axially symmetric Gaussian beams it has the following form:

$$L = \frac{n_e N_e f_e N_i}{2C_i} \cdot \sqrt{\frac{p}{p(\varepsilon_e^2+\varepsilon_i^2)+\varepsilon_e\varepsilon_i(1+p^2)}}. \quad (4)$$

Here $n_e$ is the bunch number in the electron beam, $N_i$ is ion number in the ion beam, $C_i$ is the circumference of the ion ring, $N_e$ is electron number in a bunch of the electron beam, $f_e$ is the electron revolution frequency, $\varepsilon_i$ and $\varepsilon_e$ are the emittances of the electron and ion beams. Parameter $p$ ("relative magnetic rigidity parameter") is described by the following formula:

$$p \equiv \frac{\beta_e^*}{\beta_i^*} = \frac{Z_i p_e}{p_i}, \quad (5)$$

here $\beta_{e,i}^*$ are the minimum values of the beta-functions for electrons and ions, $Z_i$ is the ion charge number, $p_e$ and $p_i$ are the electron and ion momenta.

Luminosity of an e-i collider with axially symmetric bunched electron and ion beams is calculated below according to the following formula [15]:

$$L = \frac{n_e N_e f_e N_i}{(2\pi)^2 \sigma_{se}\sigma_{si} n_i} \cdot \int_{-\infty}^{\infty} d\xi \int_{-\infty}^{\infty} d\eta \frac{\exp\left\{-\frac{1}{2}\left(\frac{\eta^2}{\sigma_{se}^2}+\left(\frac{\eta+V\xi}{\sigma_{si}}\right)^2\right)\right\}}{\varepsilon_e\beta_e^*\left[1+\left(\frac{\xi+\eta}{\beta_e^*}\right)^2\right]+\varepsilon_i\beta_i^*\left[1+\left(\frac{\xi+\eta}{\beta_i^*}\right)^2\right]}, \quad V = 1 - \frac{v_e}{v_i}. \quad (6)$$

Here $\sigma_{se}$, $\sigma_{si}$ are the longitudinal Gaussian sizes of the electron and ion bunches, $n_i$ is number of bunches in the bunched ion beam.

An example of an e-i collider with two ion beam operation modes is given in Table II. Parameters of the electron beam of the collider are chosen on the basis of the experience gained with the electron-positron collider VEPP-2000 at the Budker INP (Novosibirsk) and the ELISe project at FAIR [13].

TABLE II. Parameters of the electron-ion collider.

| Particle | $^{248}$U$^{92+}$ | | Electron |
|---|---|---|---|
| Type of the beam | Coasting | Bunched | Bunched |
| Ring circumference $C_{i,e}$, m | 18.56[*)] | | 16.0 |
| Energy, MeV/u, MeV | 300 | | 500 |
| Revolution frequency $f_{i,e}$, MHz | 10.547 | | 18.75 |
| Particle number $N_{i,e}$ | $10^3$–$10^7$ per the beam | | $5 \cdot 10^{10}$ per bunch |
| Bunch number $n_{bunch}$ | 1 | 16 | 9 |
| Bunch length $\sigma_s$, cm | — | 15 | 4 |
| Beam emittance $\varepsilon_{i,e}$ | 1 pm–100 nm | ≤ 50 nm | 50 nm |
| Beam transverse size $\sigma_x$, mcm | — | ≤ 220 | 220 |
| Laslett tune shift $\Delta Q$ | ≤ 4.4 × 10$^{-5}$ | ≤ 0.0022 | 2.7 × 10$^{-6}$ |
| Beam-beam tune shift $\xi$ | 0.07 | | ≤ 1.6 × 10$^{-4}$ |
| Minimum beta-function $\beta^*$, m | 1.0 | | 1.0 |

[*)] Ion ring circumference is chosen to meet the synchronization condition (8) in section 4

For calculation of the luminosity at the collision of a coasting ion beam with a bunched electron beam we assumed variation of the ion beam emittance according to (3). In the bunched ion beam mode the beam emittance is found from the condition of the Lasslett tune shift constancy (see section 4).

The value of $\beta^*$ is chosen to be 1 m for both ion and electron rings. It means the relative magnetic rigidity parameter (5) is equal to $p = 1$. This is possible if the magnetic focusing system of the rings is designed as proposed in the ELISe project [12] — the elements of the final focus of both rings at the interaction point are operated independently. Another way to provide independent focusing of both beams is so called "crab crossing" collision scheme (see [16,17]) where two beams collide at a large crossing angle without luminosity loss. In the case of a coasting ion and bunched electron colliding beams, the "crab crossing" bunch rotation is applied only to the latter.

The results of the luminosity calculations are presented in Fig. 2 for three versions of the ion beam:

- coasting beam (solid line),
- bunched beam where beam emittance $\varepsilon_i$ and the bunch transverse size $\sigma_x$ are calculated as a function of the ion number in the beam from the Laslett tune shift formula at $\Delta Q = 0.05$ (dashed curve),
- emittance of the bunched beam is fixed; it and the bunch transverse size have the same values as for the electron beam (Table II); the result is shown by the dotted curve.

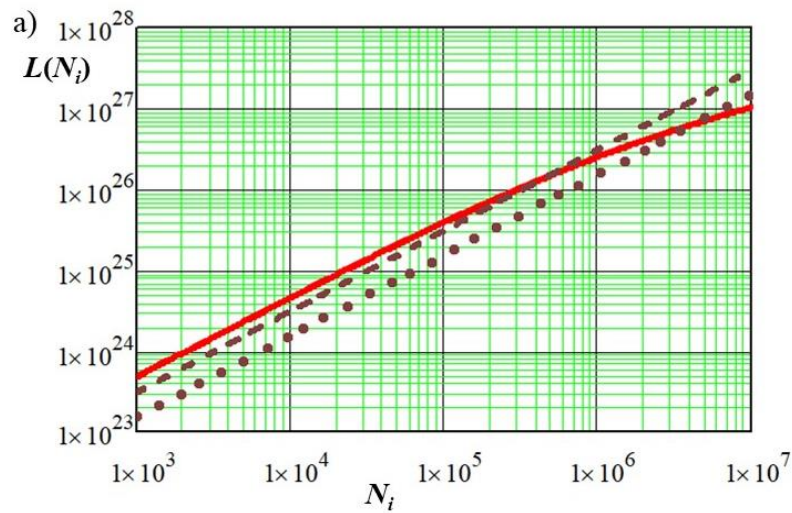

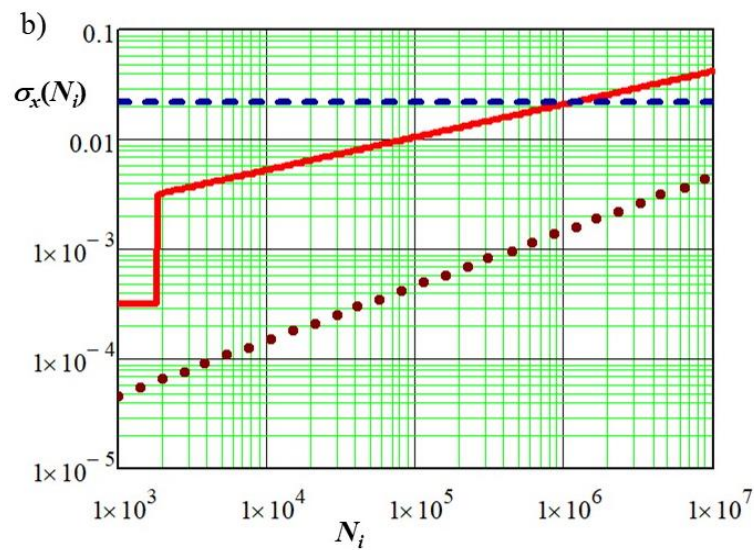

Fig. 2. Dependence on the ion number of (a) the e-i collider luminosity (cm$^{-2}\cdot$s$^{-1}$) and (b) the transverse size $\sigma_x$ (cm); solid curve — the coasting beam below and above transition, dashed curve — the bunched ion beam with variable emittance, dotted line — the bunched ion beam with a fixed emittance $\varepsilon_i = \varepsilon_e = 50$ nm.

As we see from Fig. 2a, the difference between all three versions is very small (factor 1.5–2) that has a simple explanation: the level of the luminosity for chosen parameter values is determined at $N_i < 1\cdot10^6$ by the electron beam emittance (Fig. 2b, dotted curve). It should be noted that the extremely small emittance of the bunched ion beam, calculated from the Laslett criterion, is hardly achievable in the experiment, since in the calculation the ion intrabeam scattering, which inevitably leads to an increase in the emittance, is not taken into account. However, this will not effect on the luminosity estimate (Fig. 2a, dashed curve) when the ion bunch emittance does not approach the value of the electron bunch emittance.

## 4. Collision synchronization and space charge effects

*Collision of the bunched beams.* After the collision of an ion bunch with an electron bunch the collision of the next ion (i) and electron (e) bunches occurs in vicinity of the interaction point (IP) after the shortest time

$$t_{min} = \frac{T_i}{n_i} = \frac{T_e}{n_e}. \tag{7}$$

Here $T_i$ and $T_e$ are the revolution periods of ions and electrons in their rings. During this time the next pair colliding i-e bunches reaches interaction point (IP). This equality gives us the synchronization condition:

$$f_{cool} = n_e f_e = n_i f_i. \tag{8}$$

A more general condition follows from a similar consideration: the equality

$$t_{coll} = \frac{T_i}{n_1 n_i} = \frac{T_e}{n_2 n_e} \tag{9}$$

means that now the ion bunch $n_1$ collides with the electron bunch $n_2$, where $n_{1,2}$ are the numbers of the bunches counted in the direction opposite to the bunch rotation. Condition (9) can be written as

$$f_{coll}(n_1, n_2) = n_1 n_e f_e = n_2 n_i f_i. \tag{10}$$

In the case of the collision of two bunched beams, synchronization condition (10) is met at certain values of the colliding particles' energy (velocity $v_1$, $v_2$):

$$v_i = v_e \cdot \frac{n_1 n_e C_i}{n_2 n_i C_e}. \tag{11}$$

However, the parameters $n_1$, $n_2$ are integer. Therefore, minimum variation of the particle energy is allowed with a step of $\Delta n = 1$. Scanning with a smaller step requires special variation of the particle orbit in the ring. This is especially important for colliding beams of moderately relativistic particles, which occurs, for example, in electron-ion colliders dedicated to studies in nuclear physics [12,13].

*Collisions of a bunched electron beam with a costing ion beam*

The problem of synchronization does not exist at the collision of a coasting beam with a bunched one. In this case

$$f_{coll} = n_e f_e \qquad (12)$$

because every electron bunch meets one or several ions of the coasting beam at the IP

$$(N_i)_{coll} \sim \frac{N_i}{C_i} \cdot \sigma_i.$$

Generally speaking, the colliding i-e beams can suffer from space charge effects. The most significant of them are incoherent transverse instabilities, which are described by the well-known formulas of the Laslett tune shift of particle betatron oscillations and beam-beam effect. As the estimates show (Table II), the first, Laslett, effect is negligibly small for both ion and electron beams. It results in our case from the very low intensity of the ion beam (both modes) considered here and the very large value of the relativistic factor $\gamma^3$ ($\gamma \sim 10^3$ for electrons) in the denominator of the formula for the Laslett effect. However, the second, beam-beam, effect is rather large for ions interacting with a very intense electron bunch ($\xi_{ie} \sim 0.07$). This value does not depend on the choice of the ion beam mode, coasting or bunched, because it is effect of single ion scattering in the electromagnetic field of the counterpropagating electron bunch. Therefore, if the synchronization condition for bunched beams (10) is met, this effect can have a resonant character for every bunch of the bunched ion beam and for "reference" ions of the coasting beam. The latter are the ions that "occupy" the same place in the coasting beam, as the resonant bunches in the bunched beam.

However, one can avoid such a resonance and diminish beam-beam effect for ions by choosing the parameter values — the ring circumferences and the energy — such that the reference particle collides with an electron bunch when completes a significant number of turns $N_{ti}$ after the previous collision. It can be achieved by violation of synchronization condition (10). Slightly shifting the kinetic energy of the reference ion $\varepsilon_i^0$ by factor $\delta\varepsilon_i$ one can shift its revolution frequency by $\delta f_i$:

$$f_i = f_i^0 + \delta f_i, \quad \delta f_i = \frac{\delta \varepsilon_i}{\beta_i \gamma_i^3 mcC_i} \ll f_i^0. \qquad (13)$$

Here $f_i^0$ is the ion revolution frequency that meets the condition (10) corresponding to the ion energy $\varepsilon_0$, the parameters $\beta_i$ and $\gamma_i$ are the Lorenz factors of the reference ion, $m$ is its mass, $c$ is the speed of light. For superrelativistic electrons, varying their revolution frequency $f_e$ by changing the electron energy is certainly ineffective.

Let us consider a simplified case where the first collision of the reference ion with the electron bunch occurs at the collider interaction point (IP), where the bunch center is located at this moment.

During the time $T_i^0 = (f_i^0)^{-1}$, the ion rotating in the ring with the frequency $f_i$ passes the azimuthal phase sector

$$\Delta\varphi_i = 2\pi \cdot f_i T_i^0 = 2\pi + \delta\varphi_i, \quad \delta\varphi_i = 2\pi \frac{\delta f_i}{f_i^0}. \qquad (14)$$

To avoid a parasitic collision of the reference ion with the electron bunch, the phase shift $\delta\varphi_i$ has to exceed the phase duration of the electron bunch at the IP in the ion ring

$$\delta\varphi_i \geq 2\pi \cdot \frac{2\sigma_{se}}{C_i}. \tag{15}$$

After $N_{ti}$ turns the ion returns at the IP if

$$N_{ti} = \frac{2\pi}{\delta\varphi_i} = \frac{f_i^0}{\delta f_i}. \tag{16}$$

To meet the same electron bunch there, the latter has to make $N_{te}$ turns in its own ring

$$N_{te} = f_e \cdot T_i^0 \cdot N_{ti}. \tag{17}$$

Taking into account the equality (7), we can write from (17)

$$N_{te} = \frac{n_i}{n_e} \cdot N_{ti}. \tag{18}$$

We emphasize that both $N_{ti}$ and $N_{te}$ are to be integer. This requirement is obviously met if $N_{ti}$ is a multiple of $n_e$.

For the parameters of the electron-ion collider given as an example above (Table II) at $\delta f_i/f_i^0 = 1/72$ we find

$$N_{ti} = 72, \ N_{te} = 128.$$

This ion frequency shift corresponds to the energy shift

$$\delta\varepsilon_i = \beta_i^2 \gamma_i^3 mc^2 \cdot \frac{\delta f_i}{f_i^0} = 11.875 \ MeV/u.$$

## 5. Conclusion

The main advantage of using quasi-ordered ion beams is the possibility of obtaining a luminosity sufficient for studying the properties of rare and exotic isotopes at a very low intensity of their beams.

The main critical parameter of such a beam is its linear density

$$\left(\frac{dN_i}{ds}\right)_{max} \sim 5 \cdot 10^5 \ \text{ions/m}, \tag{19}$$

when the dependence of the beam emittance on the ion energy (3) can be maintained.

Formulation of criterion (19) is based on the experimental data obtained on electron-cooled storage rings at nonrelativistic energies of ions to be cooled (Table I). At relativistic energies the value $(dN_i/ds)_{max}$ in (19) must be transformed into the system moving at an average velocity of the cooled ions.

Criterion (19) works for both coasting and bunched quasi-ordered ions beams. The use of the latter makes sense in the case of the extremely limited intensity of the ion beam:

$$N_i < \left(\frac{dN_i}{ds}\right)_{max} \cdot C_i \ .$$

However, bunching can destroy the ordering of the beam. This question needs a further study.

The possibility of detuning synchronization conditions (8, 10) when using a coasting ion beam is very important, because it reduces the resonant character of the beam-beam effect. No less important is also the possibility of the precise scanning of the ion energy while simultaneously maintaining the collider luminosity level.

**Acknowledgements**

The Author is sincerely grateful to Yu. Oganessian and Boris Sharkov for initiation and support of this work, M. Steck for providing plenty of experimental information and discussion of physics of ordered ion beams, P. Shatunov and Yu. Shatunov, L. Grigorenko, A. Fomichev, V. Lebedev, and S. Nagaitsev for very fruitful discussions of electron-ion collider issues.

**REFERENCES**

[1] E. N. Dementev, N. S. Dikansky, A. S. Medvedko, V. V. Parkhomchuk and D. V. Pestrikov, Measurement of the thermal noise of a proton beam on NAP-M storage ring, J. of Technical Physics, 50 (1980) pp. 1717–1721; Preprint INP SB USSR 79–70 (1979) Novosibirsk; Preprint CERN PS/AA 79–41 (1979) Geneva.
[2] V. V. Parkhomchuk, Study of fast electron cooling, Proceedings of the Workshop on Electron Cooling and Related Applications, 1984, edited by H. Poth, KfK Report No. 3846 (1985) p. 71.
[3] M. Steck et al., Anomalous temperature reduction of electron-cooled heavy ion beams in the storage ring ESR, Phys. Rev. Lett. 77 (1996) p. 3803.
[4] M. Steck et al., New evidence for one-dimensional ordering in fast heavy ion beams, J. Phys. B 36 (2003) pp. 991–1002; NIM A 532 (2004) p. 357.
[5] M. Steck et al., Ordered ion beams, Proceedings of EPAC 2000, Vienna, Austria, pp. 274–276.
[6] H. Danared et al., One-dimensional ordering in coasting and bunched beams, J. Phys. B 36 (2003) pp. 1003–1010.
[7] T. Shirai, A. Noda, I. Meshkov, A. Smirnov et al., Electron cooling experiments at S-LSR, Proceedings of COOL 2007, Bad Kreuznach, Germany, THM1I02 (2007) pp. 139–143.
[8] T. Shirai, A. Noda, I. Meshkov, A. Smirnov et al., One-dimensional beam ordering of protons in a storage ring, Phys. Rev. Lett. 98 (2007) p. 204801.
[9] M. Steck, Electron cooling of heavy ions, Proceedings of Cooling Workshop in Tokyo and Mt. Fuji 2016, pp. 91–103.
[10] I. Meshkov, A. Sidorin, A. Smirnov, E. Syresin, T. Katayama, Ordered State of Ion Beams, Preprint ISSN 1344-3887, RIKEN-AF-AC-34 (2002).
[11] I. Meshkov, D. Mohl, T. Katayama, A. Sidorin, A. Smirnov, E. Syresin, G. Trubnikov, H. Tsutsui, Numerical simulation of crystalline beam in storage ring, NIM A 532 (2004) pp. 376–381.
[12] A. N. Antonov, M.K. Gaidarov, M.V. Ivanov et al., The electronion scattering experiment ELISe at the International facility for antiproton and ion research (FAIR) — a conceptual design study, NIM A 637 (2011) 60–76.
[13] L. V. Grigorenko, Project DERICA: Dubna electron-radioactive isotope collider facility, Phys. Part. Nucl. Lett. 15 N7 (2018) (in print).
[14] J. Wei, H. Okamoto, A. Sessler, Necessary conditions for attaining a crystalline beam, Phys. Rev. Lett. 80 (1998) 2606.
[15] I. Meshkov, Luminosity of a collider with asymmetric beams, Phys. Part. Nucl. Lett. 15 N7 (2018) (in print); arXiv: 1802.08447v2 [physics.acc-ph].
[16] R. Palmer, Energy scaling, crab crossing and the pair problem, Report No. SLAC-PUB-4707 (1988).
[17] K. Oide and K. Yokoya, Beam-beam collision scheme for storage-ring colliders, Phys. Rev. A 40 (1989) p. 315.